\documentclass[12pt]{article}		

\usepackage[super,sort,comma]{natbib}

\usepackage{fancyhdr}		

\usepackage[section]{placeins}   %

\usepackage{graphicx}

\usepackage{xcolor}

\usepackage{amsfonts}

\usepackage{amsmath}

\usepackage{siunitx}

\usepackage{multirow}

\usepackage{makecell}

\usepackage{hyperref}
\hypersetup{ colorlinks,
    citecolor=blue,
    filecolor=blue,
    linkcolor=blue,
    urlcolor=blue
}

\usepackage[all]{hypcap}    

\makeatletter \renewcommand\@biblabel[1]{$^{#1}$} \makeatother
\newcommand{\captionv}[3]{\begin{center}\parbox{#1}{\caption[#2]{{\sf #3}}}
        \end{center}}

\newcommand{\cen}[1]{\begin{center} #1 \end{center}}

\definecolor{gray}{rgb}{0.6,0.6,0.6}
\definecolor{red}{rgb}{0.85,0,0}
\definecolor{green}{rgb}{0,0.85,0}
\definecolor{blue}{rgb}{0,0,0.85}
\definecolor{beige}{rgb}{0.92,0.87,0.78}

\newcommand{\ie}{{\it i.e.}, }
\newcommand{\eg}{{\it e.g.}, }
\newcommand{\etal}{{\it et al}}

\newcommand{\ug}{\si{\micro\gram}{}}
\newcommand{\um}{\si{\micro\meter}{} }
\newcommand{\nm}{\si{\nano\meter}{} }

\newcommand{\ums}{\si{\micro\meter}{}}

\newcommand{\cms}{\si{\centi\meter}{}}

\newcommand{\peri}{``peri''{} }
\newcommand{\oendo}{``1-endo''{} }
\newcommand{\fendo}{``4-endo''{} }
\newcommand{\peris}{``peri''{}}
\newcommand{\oendos}{``1-endo''{}}
\newcommand{\fendos}{``4-endo''{}}

\newcommand{\cellDEF}{n,cDEF{} }
\newcommand{\cellDEFs}{n,cDEF{}}

\begin{document}

\cen{\sf {\Large {\bfseries Multiscale Monte Carlo simulations of gold nanoparticle dose-enhanced radiotherapy I. Cellular dose enhancement in microscopic models} \\
\vspace*{10mm}
Martin P. Martinov, Elizabeth M. Fletcher, and Rowan M. Thomson$^\text{a)}$} \\ 
Carleton Laboratory for Radiotherapy Physics, Department of Physics, Carleton University, Ottawa, Ontario, K1S 5B6, Canada \\ \vspace{5mm} 
submitted on \today \\
}

\setcounter{page}{1}
\pagestyle{plain}
email: a) rthomson@physics.carleton.ca \\ 

\begin{abstract}

\noindent\textbf{Background:} 
The introduction of Gold NanoParticles (GNPs) in radiotherapy treatments necessitates considerations such as GNP size, location, and quantity, as well as patient geometry and beam quality.  Physics considerations span length scales across many orders of magnitude (nanometer-to-centimeter), presenting challenges that often limit the scope of dosimetric studies to either micro- or macroscopic scales. 

\noindent\textbf{Purpose:} To investigate GNP dose-enhanced radiation Therapy (GNPT) through Monte Carlo (MC) simulations that bridge micro-to-macroscopic scales.  The work is presented in two parts, with Part I (this work) investigating accurate and efficient MC modeling at the single cell level to calculate nucleus and cytoplasm Dose Enhancement Factors (\cellDEFs{}s), considering a broad parameter space including GNP concentration, GNP intracellular distribution, cell size, and incident photon energy.  Part II then evaluates cell dose enhancement factors across macroscopic (tumor) length scales.

\noindent\textbf{Methods:} Different methods of modeling gold within cells are compared, from a contiguous volume of either pure gold or gold-tissue mixture to discrete GNPs in a hexagonal close-packed lattice.  MC simulations with EGSnrc are performed to calculate \cellDEFs{} for a cell with radius $r_\text{cell}=7.35$~\um and nucleus $r_\text{nuc} = 5$~\um considering 10~to 370~keV incident photons, gold concentrations from 4 to 24~mg$_\text{Au}$/g$_\text{tissue}$, and three different GNP configurations within the cell: GNPs distributed around the surface of the nucleus (perinuclear) or GNPs packed into one (or four) endosome(s).  Select simulations are extended to cells with different cell (and nucleus) sizes: 5~\um (2, 3, and 4~\ums), 7.35~\um (4 and 6~\ums), and 10~\um (7, 8, and 9~\ums). 

\noindent\textbf{Results:} \cellDEFs{}s are sensitive to the method of modeling gold in the cell, with differences of  up to 17\% observed; the hexagonal lattice of GNPs is chosen (as the most realistic model) for all subsequent simulations.  Across cell/nucleus radii, source energies, and gold concentrations, both nDEF and cDEF are highest for GNPs in the perinuclear configuration, compared with GNPs in one (or four) endosome(s).
Across all simulations of the ($r_\text{cell}$, $r_\text{nuc}$) = (7.35, 5)~\um cell, nDEFs and cDEFs range from unity to 6.83 and 3.87, respectively.  Including different cell sizes, nDEFs and cDEFs as high as 21.5 and 5.5, respectively, are observed.  Both nDEF and cDEF are maximized at photon energies between 10 and 20~keV above the K- or L-edges of gold.

\noindent\textbf{Conclusions:} Considering 5000 unique simulation scenarios, this work comprehensively demonstrates many physics trends on DEFs at the cellular level, including demonstrating that cellular DEFs are sensitive to gold modeling approach, intracellular GNP configuration, cell/nucleus size, gold concentration, and incident source energy. These data should prove especially useful in research as well as treatment planning, allowing one to optimize or estimate DEF using not only GNP uptake, but also account for average tumor cell size, incident photon energy, and intracellular configuration of GNPs.  Part~II will expand the investigation, taking the Part~I cell model and applying it in \cms{}-scale phantoms.

\noindent\textbf{Keywords: Gold Nanoparticles, Monte Carlo, Microdosimetry}
\end{abstract}

\setlength{\baselineskip}{0.7cm}
\pagestyle{fancy}

\section{Introduction} \label{sec_Cell:Introduction}\vspace{-2mm}

The use of Gold NanoParticles (GNPs) as radiosensitizers has been investigated with increasingly more complex \textit{in vitro} and Monte Carlo (MC) studies.  There have been proposed delivery methods to add GNPs to existing radiotherapy treatments\cite{BN17,Di20} and discussion of the viability of using GNP dose-enhanced radiation Therapy (GNPT) in the clinic\cite{He17}.  Frequently, the quantification of the effects of such proposed treatments relies on a relatively simple Dose Enhancement Factor (DEF), defined as the ratio of dose with gold to dose without gold, that scales only with fixed, local GNP concentrations and ignores fluctuations in DEF across a large mass of tissue in GNPT.  Due to their diminutive nature, GNPs interact with tissue on a cellular level and biological factors modulate GNP uptake and configuration in cells.  In addition to effects on local dose distributions, introducing GNPs to a tumor-sized volume will cause perturbations to the radiation spectrum throughout the tumor\cite{MT17}, further complicating DEF predictions.  Thus, a {\it multiscale} GNPT model, one that can handle effects of GNPs across micro- and macroscopic scales, is required \cite{MT17,ZS16}. 

Previous work to assess the impacts of GNPT on a microscopic (cellular) level demonstrated the importance of aspects such as the approach to GNP modeling, the configuration and concentration of GNPs in the cell, and cell and nucleus size.  For example, Douglass \etal\cite{Do13} used random cell modeling to determine dose variation based on the location of GNPs within the cell.  Though their work accounted for the stochastic nature of cells (varying morphology, orientation) and considered a small volume of cells, their gold model consisted of a large contiguous volume within the cell rather than discretely-modeled GNPs. Jones \etal\cite{Jo10}  performed simulations of discrete GNPs, using the dose kernel around a single GNP to extrapolate dose enhancement on a larger scale, but neglecting the shielding effects of neighboring GNPs. Zhao \etal\cite{Zh21} used a single cell model containing discrete GNPs of different sizes in various configurations to model DNA double strand breaks at two beam energies, but ignored effects of GNPs in one cell on neighboring cells.  Kirkby and Ghasroddashti\cite{KG15} assessed mitochondrial dose enhancement, investigating different source spectra and varying GNP and mitochondria dimensions using a mathematical model to combine macroscopic fluence with mitochondrial dose calculations.  These studies provide examples of calculations that consider some aspects of GNPT that are often neglected, \eg modeling gold in different positions within cells, modeling GNPs individually rather than in contiguous/homogenized volumes,  and including the presence of gold in small surrounding scatter volumes.  However, the approaches taken are generally unable to account for important physics effects at once, including considerations that span micro- to macroscopic scales.

The present work is a comprehensive investigation of the effects of GNPs on cellular dose at a tumor level, bridging micro- and macroscopic considerations through development of multiscale MC models of GNPT.  This work is split into two papers,  referred to as Part~I and Part~II.  In the present paper (Part~I), a microscopic cell model is developed and used to compute nucleus and cytoplasm DEFs. After comparing three different approaches to modeling GNPs within the cell, we adopt the most accurate approach to consider questions such as how cell DEFs vary with gold concentration, intracellular GNP configuration, incident photon energy (range: 10~-~300~keV), and cell/nucleus radius.  The second paper (Part~II) uses the results for the microscopic cell model to investigate cell DEFs on a tumor scale.  We assess intrinsic variations in DEFs due to fluctuating local GNP concentration and cell/nucleus sizes, as well as embedding microscopic cell models throughout tumor-sized phantoms (using the previously-published heterogenous multiscale (HetMS) framework\cite{MT17}) to determine subcellular DEFs throughout a treatment volume. The statistical fluctuations in DEFs with varying GNP configuration and uptake are used to generate a range of expected DEF values of a population of cells as an analog to single cell nominal DEF values. The multiscale GNPT model developed herein (Parts~I and II) can also serve as a useful tool to calculate expected cell DEFs more generally for even the most complex GNPT treatments.

\section{Methods} \label{sec_Cell:Methods}

\subsection{Monte Carlo parameters} \label{ssec:Methods-MC}

All simulations performed in this paper are based on EGSnrc commit 822ec3a of the EGSnrc GitHub branch except as noted here.  A modified version of the application egs\_chamber\cite{Wu08} (with the ability to score in multiple regions simultaneously) using the egspp class library\cite{Ka05a} with the additional egs\_lattice geometry library\cite{MT20} is used for all simulations.  Compilation is done using the full GNU compiler suite for LINUX (gfortran, gcc, g++).  Validation of this geometry was performed by Martinov and Thomson\cite{MT20}.  Application egs\_chamber is compiled with the EADL\_RELAX macro set to true which ensures atomic relaxations are modeled for all energies and transitions (which is now the default with EGSnrc).  Default parameters are used for simulations except as noted here.  NRC cross sections are used for bremsstrahlung events.  XCOM cross section data are used for photon interactions\cite{Be10}.  Transport is performed down to 1~keV photon and electron kinetic energies.  Pair angular sampling is turned off.  Electron impact ionization and Rayleigh scattering are turned on (default values). Statistical uncertainties are calculated on a history-by-history basis.  All simulations are run for $10^{11}$ histories and no variance reduction techniques are used. The single cell simulations were performed on a single thread Intel Xeon 5160 (3.00 GHz) core and took from under an hour (for high GNP concentration and higher DEFs) to at most a few hours to complete (for low concentrations and DEFs close to 1).

All media used in the cell simulations are either gold, ICRU four-component tissue (10.1\% hydrogen, 11.1\% carbon, 2.6\% nitrogen, and 76.2\% oxygen by mass\cite{ICRU44}), or a homogeneous mixture of the two.  Spherical cells are modeled with a central sphere representing the nucleus (radius $r_\text{nuc}$) and an outer spherical shell representing the cytoplasm (radius $r_\text{cell}$).  A reference cell ($r_\text{cell}$, $r_\text{nuc}$) = (7.35, 5)~\um is used for all cell models\cite{OT17a}, except in Section~\ref{ssec:Methods-Sensitivity} where different cell and nucleus radii are investigated.  Cells are placed at the center of a 100~\um radius spherical phantom containing only tissue.  Thus, the gold within the cell is the only gold in the simulation.  To compare with results for cells simulated with GNPs, an all-tissue cell with no GNPs is also simulated to compute nucleus and cytoplasm dose.  The no-gold cell doses are then used to calculate DEFs for the nucleus and cytoplasm (nDEF and cDEF, respectively; \cellDEFs{}, collectively) by dividing the dose to the nucleus or cytoplasm when gold is present by the dose to the nucleus or cytoplasm in the no-gold simulations.  Monoenergetic photons of energies ranging 10-370~keV are generated isotropically within a spherical shell at the outer edge (1~\nm from the outside) of the 100~\um radius phantom.

\subsection{GNP modeling approach within a cell} \label{ssec:Methods-Cell}

\begin{figure}[htbp]
	\centering
	\includegraphics[width=0.95\textwidth]{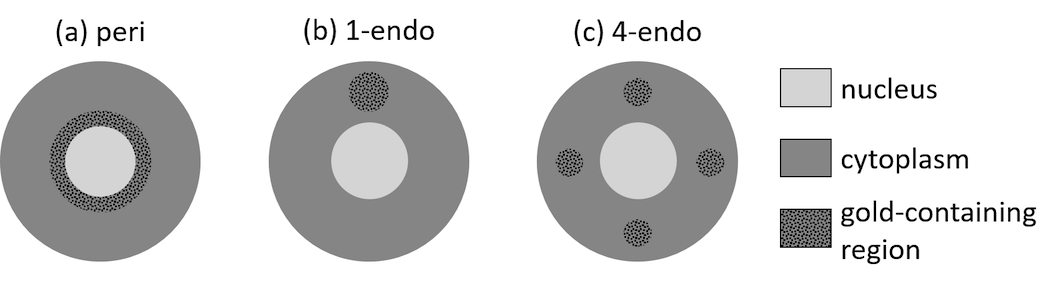}
	\captionv{0.95\linewidth}{Cell GNP configuration diagram}{A 2D cross-sectional diagram of a cell with three different gold configurations: (a) perinuclear, (b) single endosome, and (c) four endosomes.  All endosomes in (c) are translated into the same plane for visualization.
	\label{fig:GoldConfig}}
\end{figure}

Three gold configurations are considered by defining different volumes within a cell as the `gold-containing region', shown in Fig.~\ref{fig:GoldConfig}.  The first gold-containing region is a thin shell around the nucleus, called the perinuclear (\peris) configuration, which represents gold accumulating in the region close to the cell nucleus\cite{Ch10c}.  The second configuration is a region which consists of a sphere placed with its center equidistant to the nuclear radius and outer radius of the cell, called the one endosome (\oendos) case, which represents GNPs aggregating in a single compartment which sits in the cytoplasm\cite{CC07}.  The final configuration is similar to the \oendo configuration, but with the volume split evenly amongst four spheres placed at the four vertices of a tetrahedron, called the four endosomes (\fendos) case. The four endosome centers are also equidistant to the nucleus and outer radius of the cell. The scoring volumes are the entire nucleus and the region of the cytoplasm that does not contain gold, to compute nDEF and cDEF respectively. Regardless of gold configuration, for a given concentration each cell (of the same size) contains the same mass of gold. 

GNPs are modeled as 25~\nm radius spheres of pure gold (when modeled explicitly).  A minimum 5~\nm buffer region in which no dose is scored is added around each gold-containing region (with a minimum 5~\nm space between GNPs when modeled explicitly) for all cases; this buffer accounts for some of the polyethylene glycol (PEG) surface coatings used in GNP delivery\cite{KK17}. Typical gold concentrations investigated are 5, 10, and 20~mg$_\text{Au}$/g$_\text{tissue}$ (abbreviated as mg/g), as well as all integer concentrations between 4 and 24~mg/g for a subset of energies. These concentrations are consistent with previous MC studies (\eg see Refs.~\cite{KK16,Br17,Ba20}) and gold concentration is discussed in Section~\ref{sec_Cell:Discussion}. All gold is assumed to be taken up by the cell (\ie no gold in extracellular medium) and the total mass of gold per cell is chosen such that the average gold concentration of the overall medium remains the same.  For example, for an average gold concentration $c$ (units of mg/g), the number of GNPs in a cell ($n_\text{GNP}$) must satisfy
\begin{equation}
	\frac{m_\text{GNP} \cdot n_\text{GNP} \cdot N_\text{cell}}{\rho_{\text{tissue}}} = c \label{eq:conc}
\end{equation}
where $m_\text{GNP}$ is the mass  of a single GNP and $N_\text{cell}$ is number of cells per \cms$^3$ of tissue (\eg $3\times10^8$ reference cells per \cms$^3$)\cite{Th13}; $\rho_{\text{tissue}}$ is the tissue mass density which is 1~g/\cms$^3$ for the ICRU tissue used in this model.  This equation does not account for the volume of tissue displaced by GNPs, but for all concentrations studied in this work the correction had a negligible effect on calculated values.

\begin{figure}[htbp]
	\centering
	\includegraphics[width=0.4\textwidth]{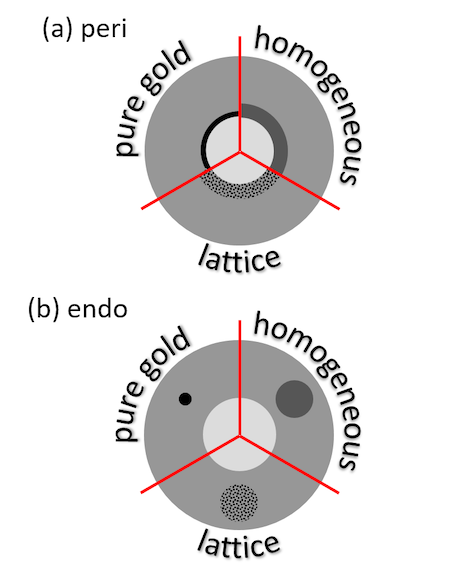}
	\captionv{0.95\linewidth}{Cell GNP model diagram}{Diagram of three approaches to modeling gold in (a) a perinuclear shell or (b) an endosome; contiguous pure gold (upper left), homogeneous gold-tissue mixture (upper right) and a lattice of discrete GNPs (bottom) for each cell image.
	\label{fig:GoldCell}}
\end{figure}

Three different approaches to modeling gold within a cell are investigated to ensure accurate but efficient modeling (Figure~\ref{fig:GoldCell}.  The gold-containing region is modeled as containing: contiguous gold, a homogeneous gold-tissue mixture, or  a hexagonal lattice of GNPs embedded in tissue.  For the hexagonal lattice, GNPs intersecting the boundary are modeled as only the partial volume of the sphere within the gold-containing region.  A contiguous gold model was used by Douglass \etal\cite{Do13}; a hexagonal prism GNP lattice model was used by Cai \etal\cite{Ca13}; the homogeneous mixture model is introduced as a third model that is as simple to implement as the contiguous gold model but still accounts for the tissue medium between GNPs.  For the \peri configuration, the gold-containing region is modeled as a spherical shell about the nucleus with a thickness of 2.7-10.9~\nm in the pure gold model and 50~\nm (diameter of a GNP) in the other two models, with GNP number density varying between $8.294\times10^{14}$ and $3.316\times10^{15}$~GNP/\cms$^3$.  In either endosome case, the gold-containing region is modeled as a sphere with a radius between 372 and 591~\nm in the pure gold model and a radius between 494 and 783~\nm for the homogeneous and lattice models.  GNP number density in the endosomes is always $6.457\times10^{15}$~GNP/\cms$^3$, the maximum density assuming 5~\nm spacing between GNPs. 

\subsection{Cell DEFs for variable cell and nucleus radii} \label{ssec:Methods-Sensitivity}

Simulations are performed for the three gold configurations in Section~\ref{ssec:Methods-Cell} (\ie \peris, \oendos, and \fendos) using the discrete GNP modeling approach for different cell sizes. Cells with radii (and corresponding nucleus radii) of 5~\um (2, 3, and 4~\ums), 7.35~\um (4, 5, and 6~\ums), and 10~\um (7, 8, and 9~\ums), chosen to represent a range of cell sizes\cite{OT17a}, are investigated for concentrations of 5, 10, and 20~mg/g.  The 5 and 10~\um radius cells are assumed to maintain the same nearest neighbor distance (2.06~\ums), thus the cell number density for the larger and smaller cells is $8.06\times10^8$ and $1.32\times10^8$~cells/\cms$^3$, respectively\cite{OT17a}, as opposed to $3\times10^8$~cells/\cms$^3$  for the reference cell\cite{Th13}.  To ensure the total gold concentration throughout the medium remains consistent, small and large cell GNP uptake is adjusted to compensate for the difference in cell to extracellular volume ratios.  For example, following equation~(\ref{eq:conc}), the number of 25~\nm radius GNPs ($n_\text{GNP}$) in a 5, 7.35 or 10~\um radius cell would be $1.96\times 10^4$, $5.27\times 10^4$, and $1.21\times 10^4$, respectively, to achieve a concentration of 20~mg/g. These numbers are consistent with the expected range for 50~nm diameter GNPs established by Zhao \etal\cite{Zh21} (see Figure~3 in that work).

\section{Results} \label{sec_Cell:Results}

\subsection{GNP modeling approach within a cell} \label{ssec:Results-Cell}

\begin{figure}[htbp]
	\centering
	\includegraphics[width=0.4\textwidth]{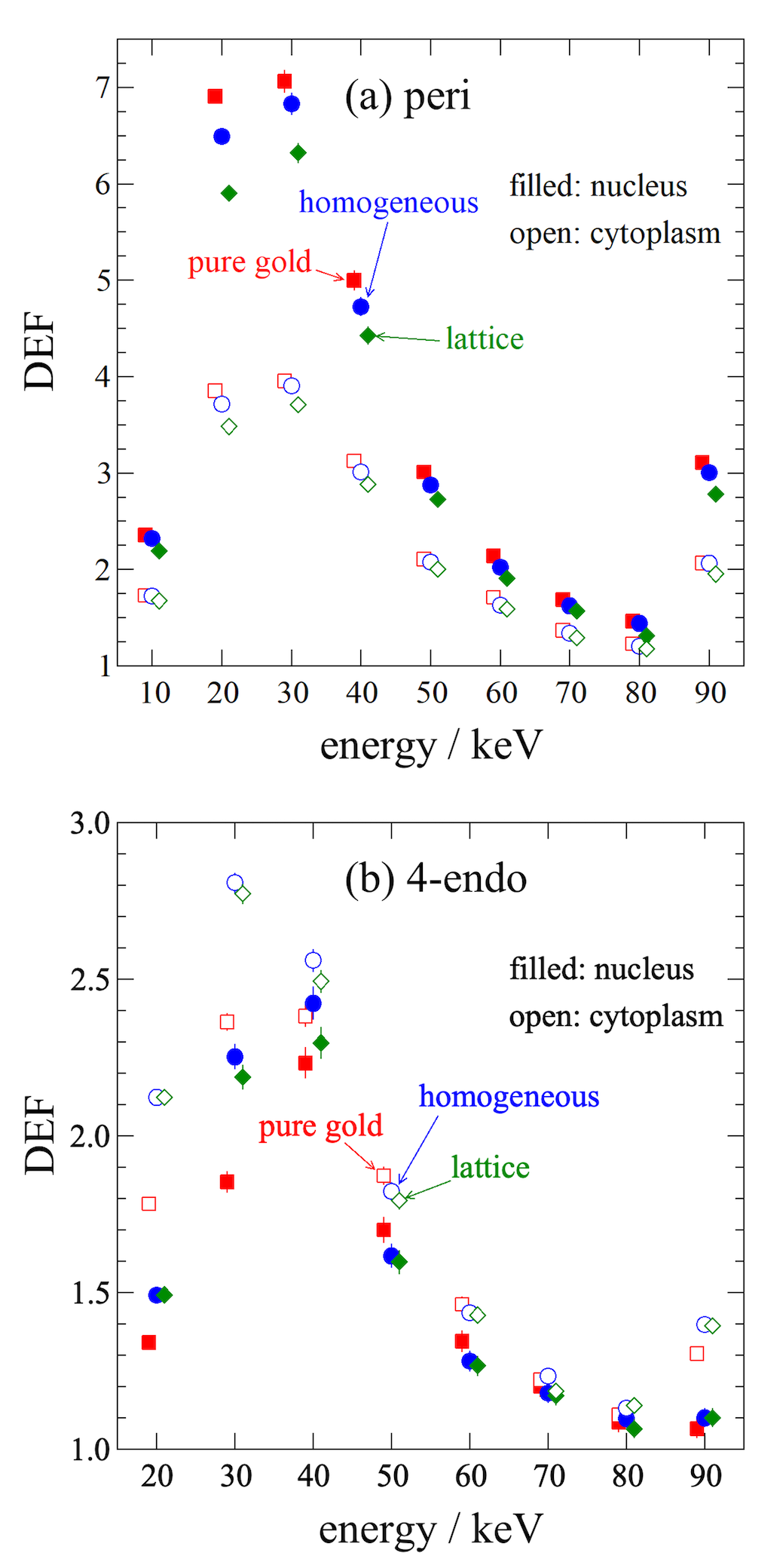}
	\captionv{0.95\linewidth}{\cellDEFs{} vs energy for different gold models and configurations}{Nucleus and cytoplasm DEF of the ($r_\text{cell}$, $r_\text{nuc}$) = (7.35, 5)~\um reference cell for the (a) perinuclear and (b) four endosome gold configurations when gold (20~mg/g concentration) is modeled using contiguous gold (red squares), a homogeneous gold-tissue mixture (blue circles), or a hexagonal lattice (green diamonds).
	\label{fig:SingleCellModels}}
\end{figure}

Figure~\ref{fig:SingleCellModels} compares DEF of the nucleus and cytoplasm of the reference cell as a function of energy for the three different methods of modeling gold described in Section~\ref{ssec:Methods-Cell}.  Results are shown for the \peri and \fendo configurations at 20~mg/g gold concentration. For the \peri configuration, both nucleus and cytoplasm DEFs (\cellDEFs{}s) are highest when gold is modeled contiguously; 3.5-11\% (cytoplasm) and 7.5-17\% (nucleus) higher than the analogous lattice model DEFs.  The homogeneous model \cellDEFs{}s are still 2.4-6.7\% (cytoplasm) and 3.7-10\% (nucleus) higher than the lattice model DEFs.  The three different modeling approaches yield different results for most DEFs above 2 for the \peri configuration. This sensitivity to modeling approach results from differences in the distance between a point in the gold-containing region and the nearest scoring volume, in the total volume of the gold-containing region, and in energy deposition to the gold and non-scoring tissue media between the different models.

For the \fendo configuration, there is good agreement between the homogeneous mixture and lattice model nDEFs and cDEFs (within 1\% for most energies), with larger differences (ranging 1.2-5.5\%) at energies above the L-edges ($\sim$13.3~keV) and K-edge (80.7~keV) of gold.  DEFs for the pure gold model are as much as 16\% lower than the lattice DEFs.  These discrepancies are due to the smaller size of pure gold endosomes (required to have consistent total mass of gold across the models); the contiguous gold endosomes have 43\% of the volume (54\% of the surface area) of the two other models.  Simulations of the pure gold case are also performed where the contiguous gold containing region occupies the same volume as the other two models (which considerably increases the total mass of gold in the cell, for which the hexagonal lattice \cellDEFs{}s were only 0.3-5.7\% higher than the pure gold model (results not shown).

The \cellDEF trends seen in Figure~\ref{fig:SingleCellModels} are qualitative agreement with those for concentrations of 5 and 10~mg/g but those have DEFs closer to unity.  The results for the \oendo configuration (for gold concentrations of 5 and 10~mg/g) exhibit the same trends as the \fendo configuration.  Though there is good agreement between different modeling approaches in many cases, the discrepancy between the three models is greatest at 20-40~keV where DEFs are higher.  Therefore, moving forward in this work, only the hexagonal lattice of discrete GNPs is used to model gold in the gold-containing regions, chosen because it is the most realistic representation of GNPs in the cell.
%
\subsection{GNP configuration and energy} \label{ssec:Results-Cell2}

\begin{figure}[htbp]
	\centering
	\includegraphics[width=0.38\textwidth]{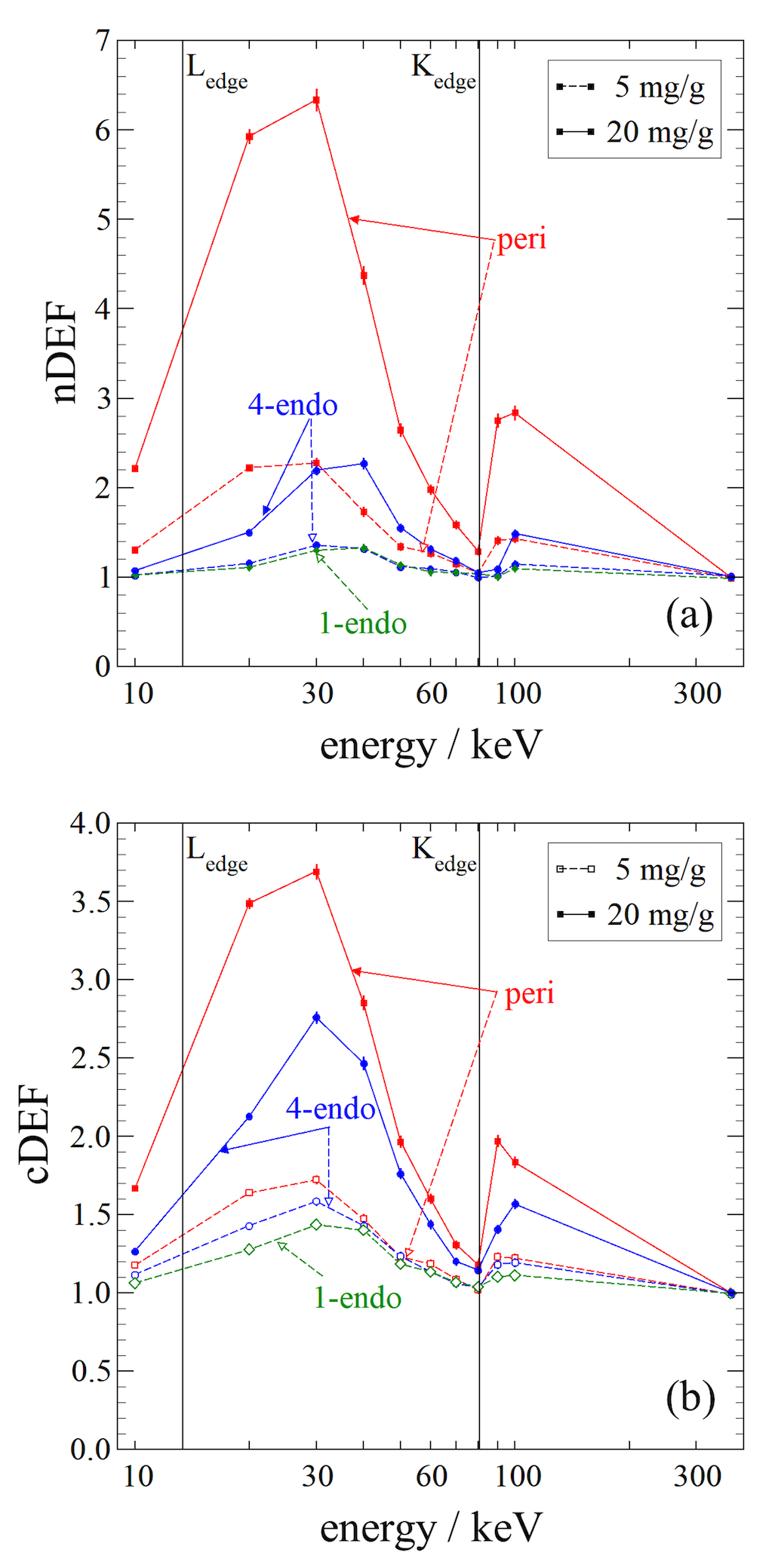}
	\captionv{0.95\linewidth}{\cellDEFs{} vs energy for different concentrations and configurations}{DEF as a function of energy for (a) nucleus and (b) cytoplasm for the ($r_\text{cell}$,~$r_\text{nuc}$)~=~(7.35, 5)~\um reference cell for three gold configurations and concentrations of 5 and 20~mg/g.
	\label{fig:SingleCellDEFs}}
\end{figure}

Figure~\ref{fig:SingleCellDEFs} expands on the data presented in Figure~\ref{fig:SingleCellModels} to present \cellDEFs{}s for energies of 10-370~keV and gold concentrations of 5 and 20~mg/g for all three gold configurations.  The largest DEFs (for any concentration or configuration) are at 20 and 30~keV, energies right above the L-edges\cite{Be10} of gold.  DEFs are also large at 90 and 100~keV (above the K-edge\cite{Be10}) but are markedly lower than at the L-edges.  The DEFs for the \oendo configuration at 20~mg/g are omitted as the required endosome size is larger than what can fit in the reference cell cytoplasm.  DEFs calculated with the \peri configuration are either equal to or higher than either endosome configuration at all energies for the same concentration.  The \fendo configuration DEFs are slightly larger than the \oendo DEFs.

\subsection{GNP configuration and gold concentration} \label{ssec:Results-Cell3}

\begin{figure}[htbp]
	\centering
	\includegraphics[width=0.8\textwidth]{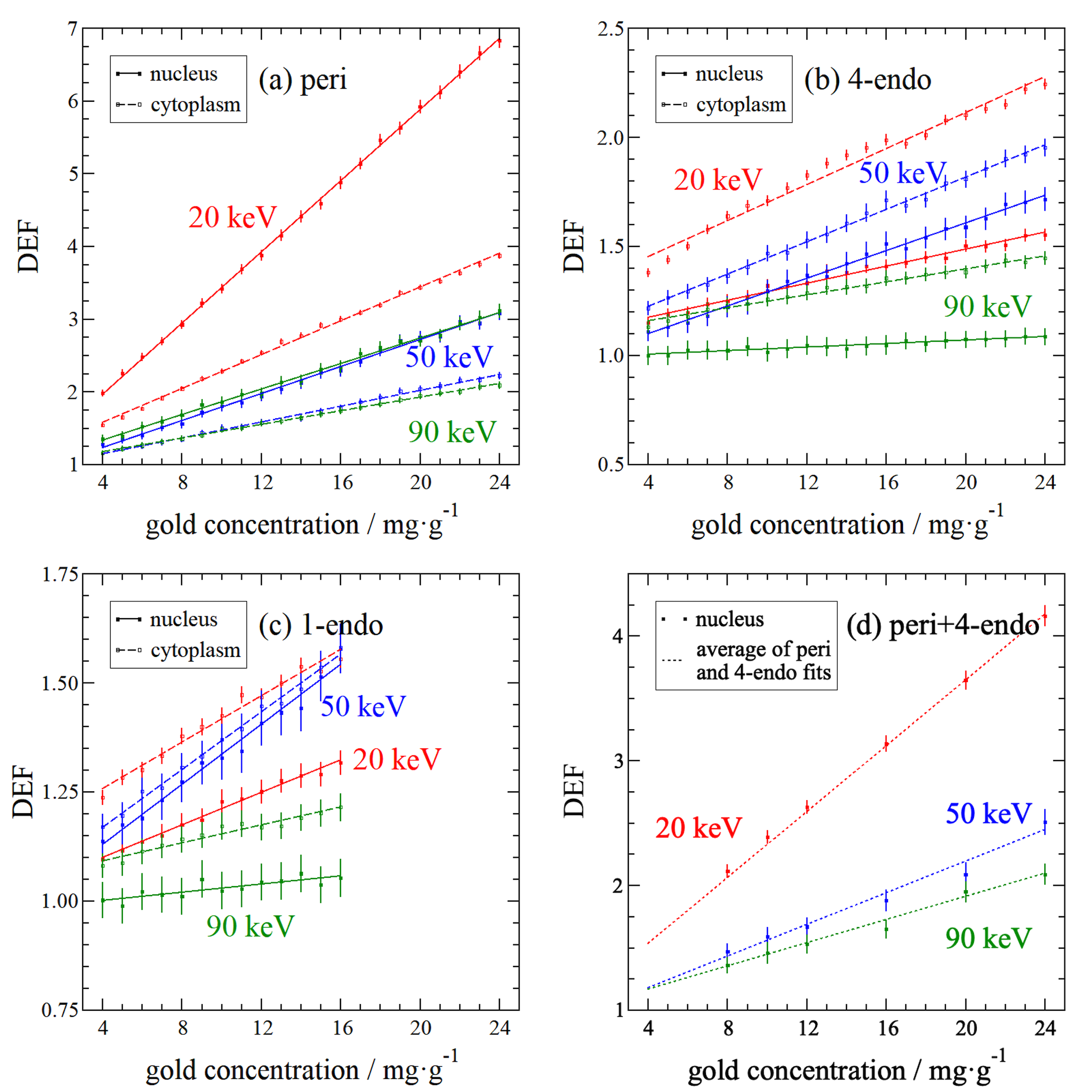}
	\captionv{0.95\linewidth}{\cellDEFs{} vs gold concentration for different energies and configurations}{Nucleus and cytoplasm DEF as a function of gold concentration in the ($r_\text{cell}$,~$r_\text{nuc}$)~=~(7.35, 5)~\um reference cell for 20, 50, and 90~keV monoenergetic photon sources for (a) \peris, (b) \fendos, and (c) \oendo configurations with linear fit (Table~\ref{tab:nDEF}), as well as (d) nucleus DEF for combined ``peri+4-endo'' gold with average of \peri and \fendo fits. \oendo results for concentrations of 17~mg/g or higher are omitted as the endosomes are too large to fit in the cytoplasm.
	\label{fig:DEFvCon}}
\end{figure}

Figure~\ref{fig:DEFvCon} presents \cellDEFs{}s of the reference cell as a function of gold concentration for the \peris, \fendos, and \oendo configurations with accompanying linear fits of the data.  As before, \peri configuration DEFs are consistently higher than the other configurations for all gold concentrations.  Endosome configuration nDEFs for 90~keV energies are close to unity for most concentrations, increasing to only $\sim$1.1 for 24~mg/g concentrations.  

The parameters of the linear least squares fits of the data in Figure~\ref{fig:DEFvCon} are compiled in Table~\ref{tab:nDEF}.  Most of the DEFs computed lie on the lines of best fit (within uncertainty); the (absolute) sums of DEF residuals range from 0.027 to 0.17.  Though the fits are good predictors for the investigated range of 4-24~mg/g, several of the intercepts in the fit are significantly above or below unity (DEF for 0~mg/g is equal to one by definition), most notably cDEF intercepts at 20~keV, suggesting these fits are not reliable at very low concentrations.

These fits can be extended to combinations of the \peris, \fendos, and \oendo configurations.  A subset of MC simulations were done of cells containing both \peri and \fendo gold in equal amounts (\ie 50\% of the GNPs were in the perinuclear configuration and 50\% were in the 4-endosome configuration). The MC determined DEFs seen in these ``peri+4-endo'' cases were then compared to the average of the fits for the two configurations from Table~\ref{tab:nDEF}. As seen in Figure~\ref{fig:DEFvCon}d, nucleus dose enhancements are accurately predicted by this average fit, with agreement within approximately 2.5\%. Cytoplasm DEFs are generally also well-predicted with agreement within approximately 8\% (not shown).

\begin{table}[htbp]
	\vspace{\normalbaselineskip}
	\centering
	\captionv{0.95\linewidth}{Linear fit coefficients for \cellDEF over gold concentration}{Linear fit parameters relating gold concentration to \cellDEFs{}s of the ($r_\text{cell}$,~$r_\text{nuc}$)~=~(7.35, 5)~\um reference cell computed using the least squares method (plotted in Figure~\ref{fig:DEFvCon}), with slope in units of g$_\text{tissue}$/mg$_\text{Au}$.  Statistical uncertainty (1$\sigma$) on the final digit(s) for slopes is $\leq$0.001 g/mg, and for intercepts is indicated in parentheses. \vspace{-0.8\normalbaselineskip}
	\label{tab:nDEF}}
	\small
	\begin{tabular}{ccrrcrrcrr}
	\hline \hline
	\multirow{2}{*}{ \makecell{energy\\ (keV)} }   & \multirow{2}{*}{ target } & \multicolumn{2}{c}{ \peri } & & \multicolumn{2}{c}{ \fendo } & & \multicolumn{2}{c}{ \oendo } \\ \cline{3-4} \cline{6-7} \cline{9-10} \vspace{-0.8\normalbaselineskip} \\
	                      &      & slope & intercept     & & slope & intercept     & & slope & intercept \\ \hline \vspace{-0.8\normalbaselineskip} \\
	\multirow{2}{*}{ 20 } & nDEF & 0.245 & 0.991 (0.019) & & 0.019 & 1.025 (0.006) & & 0.019 & 1.025 (0.006) \\
	                      & cDEF & 0.116 & 1.115 (0.019) & & 0.041 & 1.290 (0.021) & & 0.026 & 1.152 (0.013) \\ \vspace{-0.8\normalbaselineskip} \\
	\multirow{2}{*}{ 50 } & nDEF & 0.093 & 0.862 (0.022) & & 0.034 & 0.993 (0.015) & & 0.034 & 0.993 (0.015) \\
	                      & cDEF & 0.055 & 0.927 (0.014) & & 0.037 & 1.077 (0.009) & & 0.033 & 1.037 (0.008) \\ \vspace{-0.8\normalbaselineskip} \\
	\multirow{2}{*}{ 90 } & nDEF & 0.088 & 0.980 (0.020) & & 0.005 & 0.983 (0.010) & & 0.005 & 0.983 (0.010) \\
	                      & cDEF & 0.047 & 0.987 (0.007) & & 0.015 & 1.099 (0.007) & & 0.010 & 1.050 (0.008) \\ \hline \hline \vspace{-0.8\normalbaselineskip} \\
	\end{tabular}
\end{table}

\newpage
\subsection{Varying cell and nucleus radii} \label{ssec:Results-Range}

\begin{figure}[htbp]
	\centering
	\includegraphics[width=0.8\textwidth]{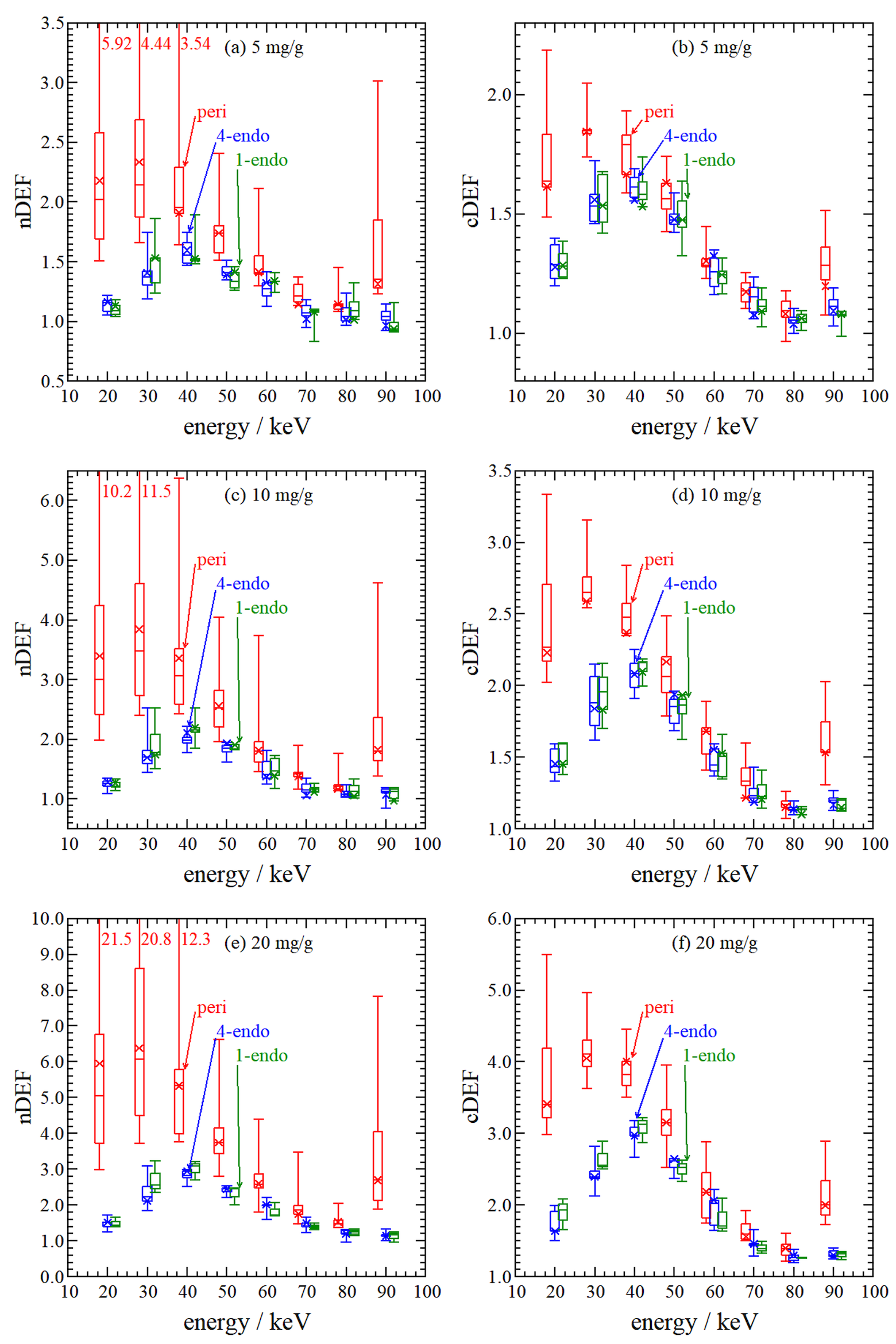}
	\captionv{0.95\linewidth}{\cellDEFs{} box and whisker plots of cell size statistics}{Box and whisker plots showing \cellDEF as a function of energy for (up to) 9 different cell/nucleus radii for gold concentrations of (a,b) 5~mg/g, (c,d) 10~mg/g, and (e,f) 20~mg/g.  The whiskers represent the total range of DEFs amongst the 9 (or fewer) permutations, the boxes represent the inner quartiles (25-75\%), the horizontal line is the median and the `x' represents the ($r_\text{cell}$, $r_\text{nuc}$) = (7.35, 5)~\um reference cell DEF.
	\label{fig:SenseWhisker}}
\end{figure}

Figure~\ref{fig:SenseWhisker}shows the range of nucleus and cytoplasm DEFs for nine different cell sizes: 
($r_\text{cell}$, $r_\text{nuc}$) = (5, 2-4), (7.35, 4-6), (10, 7-9)~\ums. 
For both nucleus and cytoplasm, the highest DEFs correspond to the (5, 2)~\um cell in most cases and the lowest DEFs correspond to the (10, 7)~\um cell.  For most \oendo (and some \fendos) configurations, cells which could not fit the size of endosome required within the cytoplasm are omitted.  For the \oendo configuration at 20~mg/g concentration, only 3 cells make up the total data range and the reference cell is among the omitted.  

Similar trends to those in Figure~\ref{fig:SingleCellModels} can be seen across all size permutations.  The largest range of DEFs over cell/nucleus size occur at energies and with gold configurations for which the reference cell DEFs are highest, \eg nDEF can range over as much as 2.99 to 21.5 in the \peri configuration at 20~keV for 20~mg/g.  The smallest ranges of DEFs occur about small reference cell DEFs, \eg cDEF ranges over 1.20 to 1.38 in the \fendo configurations at 80~keV for 20~mg/g.  At concentrations of 20~mg/g, the \oendo DEF ranges are often even smaller than the equivalent \fendo ranges; this is primarily due to the exclusion of cell sizes where the endosome cannot fit, lowering the overall number of DEFs used to populate the range.

To omit the typically higher outliers, the range of the inner quartiles (25-75\%) can be used as a more conservative estimate of DEF ranges while maintaining a lower limit relatively close to that of the full population. This range could prove clinically relevant for what one should expect in GNPT for tumor treatment planning (as overdosing by a factor of 2 or more is not typically a concern), but may be a dangerous underestimate of DEF if used for non-tumor volumes such as organs-at-risk.  The aforementioned 2.99 to 21.49 total nDEF variation is only 3.72 to 6.76 across the two inner quartiles (excluding the highest and lowest 25\% of doses).  The DEF of the reference cell (which has the median cell and nucleus size) in Figure~\ref{fig:SingleCellModels} is not the median DEF in most cases.  The reference \cellDEFs{}s are typically within the inner quartiles close to the median, though some exceptions occur particularly in scenarios with cell sizes excluded from the (statistical) population because the endosome is too large to fit in the cytoplasm.

\begin{figure}[htbp]
	\centering
	\includegraphics[width=0.4\textwidth]{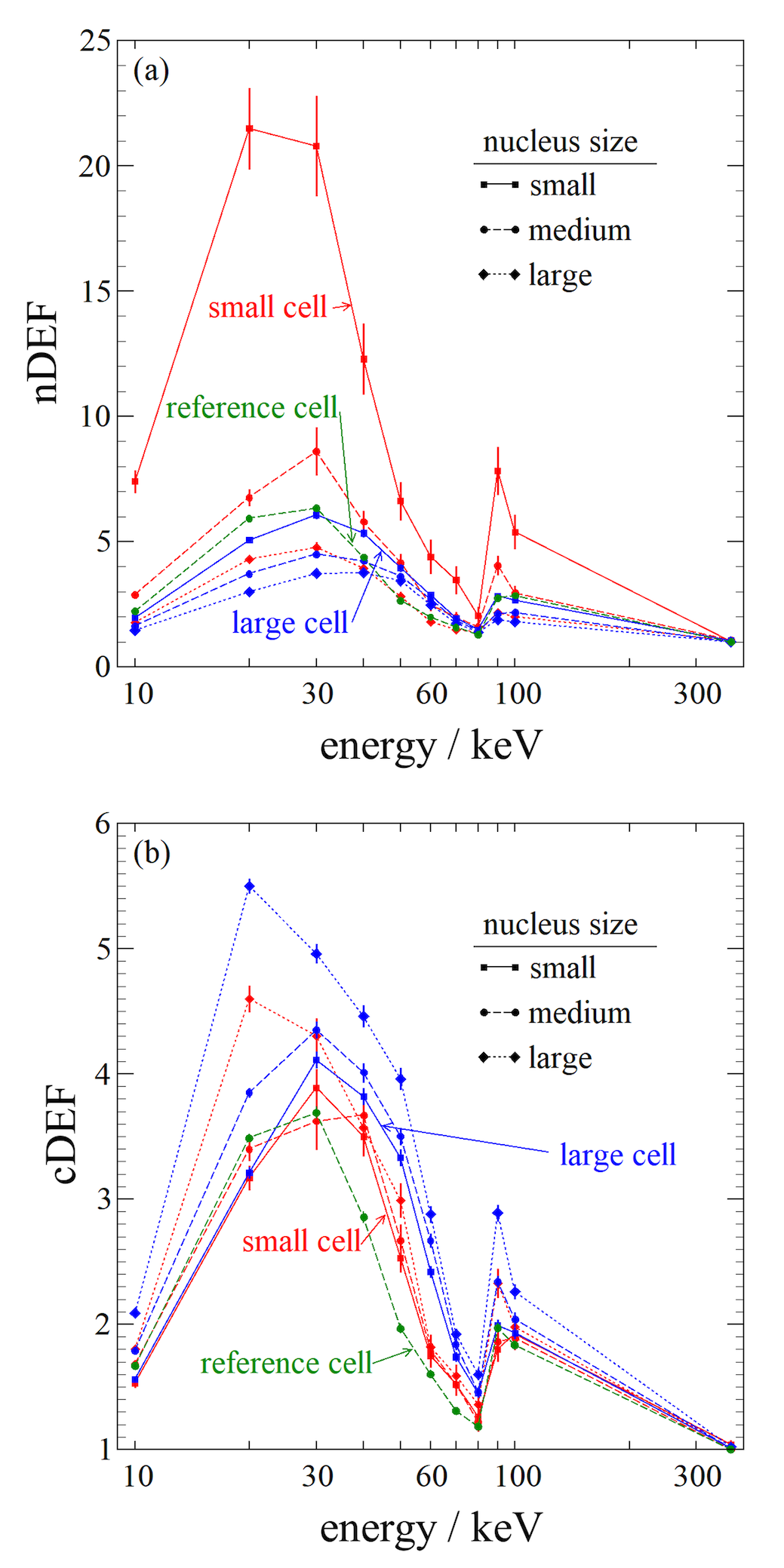}
	\captionv{0.95\linewidth}{\cellDEFs{} vs energy for different cell sizes}{DEFs as a function of energy for (a) nucleus and (b) cytoplasm of the (red) small cells, ($r_\text{cell}$, $r_\text{nuc}$) = (5, 2), (5, 3), (5, 4) \ums, and the (blue) large cells, (10, 7), (10, 8), and (10, 9)~\ums, as well as the (green) reference cell, (7.35, 5)~\ums, for 20~mg/g and gold in the \peri configuration.
	\label{fig:CellSizeDEFs}}
\end{figure}

\newpage
Figure~\ref{fig:CellSizeDEFs} plots \cellDEF as a function of energy (analogous to Figure~\ref{fig:SingleCellDEFs}) for the small (red, 5~\ums) and large (blue, 10~\ums) cells with all three of their corresponding nucleus sizes, along with the reference \cellDEF (green) for comparison.  Data are shown for GNPs in a \peri configuration with a concentration of 20~mg/g, the scenario in which DEFs are highest.  The figure shows that nDEFs (for all energies) increase as nucleus size decreases and cDEFs decrease as nucleus size decreases.  The general trends (as a function of energy) exhibited are the same for all cell sizes, but at 20 and 90~keV (10~keV above the L- and K-edges, respectively) there is a much larger (relative) increase in DEFs for cells with smaller scoring volumes, \ie high nDEFs when the nucleus is relatively small and high cDEFs when the nucleus is large (therefore cytoplasm volume is small).

\section{Discussion} \label{sec_Cell:Discussion}

This work establishes the importance of accurate modeling in simulations of GNPT. By comparing different approaches to GNP modeling within cells we show up to a 17\% difference in DEFs between the models explored here, thus demonstrating that GNPs must be modeled discretely to achieve the most accurate results. In addition, the differences in DEFs seen when varying GNP configuration within cells show that GNP position plays an important part in both nDEF and cDEF. Similarly, both  nDEF and cDEF have a strong dependence on cell and nucleus radii.

Differences seen between the three approaches to GNP modeling within the cell (Section~\ref{ssec:Results-Cell}) can be attributed to physics arising from physical differences between the three models.  For the \peri scenario there are many effects that can cause the (up to) 17\% difference between modeling approaches.  The contiguous and homogeneous models do not have gaps of pure tissue with no gold in the gold-containing region, thus photons entering the gold-containing region are guaranteed to travel through \textit{some} gold.  Additionally, photoelectrons generated in the contiguous and discrete model need to travel through some amount of pure gold (density of 19.32~g/\cms$^3$), either some portion of a 3-11~\nm thick gold shell or one (or more) 50~\nm diameter GNPs, whereas photoelectrons generated in the homogeneous model do not travel in gold, but rather the mixture (density of 1.005-1.019~g/\cms$^3$).  This will impact the energy loss of the photoelectrons before they potentially reach the scoring regions (or even the local GNP gold volume\cite{Le11c}), affecting the overall DEFs.  Compared to the \peri gold modeling considerations, the effects of using different models are less important for the \oendo and \fendo cases.  In the endosome cases, photoelectrons which are not generated in the periphery of the endosome will be mostly self-absorbed.  Thus, many of the nanostructure details which were important in the 50~\nm (or less) thick perinuclear shell average out over the (relatively) large micron-scale endosomes.  

When comparing the speeds of the simulations, the contiguous gold and homogeneous gold/tissue cell models had approximately the same efficiency.  Modeling GNPs in a discrete lattice reduced efficiency by 11 to 24\% for transport  exclusively within the cell (no extracellular regions), which, in simulations with extracellular volume, would be a smaller decrease in overall efficiency. As the discrete model utilizes the most realistic representation of GNPs in cells across all parameters at only a small efficiency cost and is recommended for general use in MC cell modeling.  

When comparing different gold configurations within the reference cell, \peri \cellDEFs{}s are as much as 3 times larger than or equal to \oendo and \fendo \cellDEFs{}s for the same gold concentration and energy; this trend is also seen when cell and nucleus radii are varied.  There are several effects at play in the \peri model causing this increase: the perinuclear region is in close proximity to both the nucleus and cytoplasm scoring volumes, the endosomes self-absorb many of the photoelectrons not generated towards their peripheries\cite{Ki17}, and in the \peri configuration the gold-containing region occupies a larger volume.  Although there are some discrepancies when comparing cDEFs between the \oendo and \fendo configurations, such as in Table~\ref{tab:nDEF} and Figure~\ref{fig:DEFvCon}, for any one energy and concentration, the cDEF discrepancy is not much larger than the statistical uncertainty of either cDEF.  Thus, because of the small discrepancies and the inability to fit a single endosome in a cell for a 20~mg/g concentration using our model, many of the \oendo results were omitted.

As shown in Figure~\ref{fig:DEFvCon}, the relationship between an individual (\ie a single cell with little surrounding scatter medium) cell DEF and gold concentration is effectively linear as the uncertainties on the slope parameters are sub-0.001~g/mg for concentrations ranging 4-24~mg/g.  Therefore, in a local model of a single cell it is possible to calculate DEF as a function of gold concentration for different cell sizes, gold configurations, and energies.  This is primarily due to the cell model constraining the gold-containing region to be the same for the \peri configuration and to be in a fixed location in the endosome configurations. The linear fits shown in Figure~\ref{fig:DEFvCon} and in Table~\ref{tab:nDEF} can also be extended to combinations of the GNP configurations studied in this work. When cells are simulated with gold in the ``peri+4-endo'' configuration (GNPs distributed equally between \peri and \fendo configurations), the nuclear DEF is well predicted by the average of the \peri and \fendo fits, with agreement within approximately 2.5\%. Cytoplasm DEF is predicted slightly less well, with agreement between data and average fit within approximately 8\%. This provides an application of the data presented here, where DEFs for combinations of configurations can be predicted based on the DEFs of the individual configurations. Unfortunately, expanding the scenario to a tumor-sized volume cannot be done directly as the difference in attenuating effects throughout the scatter media must also be taken into account due to their effects on local DEF\cite{MT17}.  

The gold concentrations considered in this work (4-24~mg/g) were motivated by previously-published research. A frequently cited GNP concentration is 7 mg/g\cite{MT17,Br20,Ch09f,Jo10}, which is well within the range of concentrations considered in this work. Further examples from MC studies include Bahreyni Toossi \etal\cite{Ba12c} who studied 10, 20, and 30~mg/g, Koger and Kirkby\cite{KK16} who studied 5-30~mg/g, Brivio \etal\cite{Br17} who studied 10-66~mg/g, Bassiri \etal\cite{Ba20} who studied 7, 18, and 50 mg/g, and Moradi \etal\cite{Mo21} who studied 5-20~mg/g. For \textit{in vitro} and \textit{in vivo} studies, observed tumor gold concentrations also tend to vary, and depend strongly on GNP size, shape, any attached ligands, and more\cite{Ch06,Ba19}. It is also important to note that \textit{in vivo} uptake to tumor (and other organs) can vary significantly from the injected GNP concentration. Zhao \etal\cite{Zh21} established a range of approximately 10,000-50,000 GNPs per cell for 50~nm radius GNPs based on the \textit{in vitro} works of Chithrani \etal\cite{Ch10b} and Peretz \etal\cite{Pe12b}; this is consistent with our range of 10,540-63,240 GNPs per cell for the nominal cell size (Eq.~\ref{eq:conc}).  Hainfeld \etal\cite{Ha13} demonstrated tumor uptake of up to 15~mg/g after the injection of 11~nm GNPs at 4~g/kg mouse bodyweight. Kimm \etal\cite{Ki20} demonstrated tumor uptake of 4.4~\ug{}/mm$^3$ (approx. equal to mg/g) after injecting mice with 5~mg of 30~nm GNPs in two doses. \textit{In vitro},  Lechtman \etal\cite{Le13} saw cellular uptake of 0.84~mg/mL after incubating cells in a 2~mg/mL solution of 30~nm GNPs. These differences in cellular/tumor uptakes, especially with advances in tumor targeting ligands, allow for gold concentrations in tumor cells to be higher than the lethal amount of gold, which is estimated to be approximately 5~mg/g\cite{Ha13,Ka16}. In our work, select concentrations of 1~mg/g and less were also investigated and were found to not follow the same linear trend for the concentrations of 4~mg/g and above, which is why several of the lines defined in Table~\ref{tab:nDEF} do not have an intercept of one; the results are omitted as computing enough DEFs to demonstrate trends at concentrations below 1~mg/g proved computationally prohibitive due to the small uncertainties needed to quantify the very small dose enhancement. Future expansion of the datasets presented here to these lower concentrations will enable comparisons to more \textit{in vivo} and \textit{in vitro} results.

In Figures~\ref{fig:SenseWhisker} and~\ref{fig:CellSizeDEFs}, the DEF trends with energy for various cell sizes are similar to those of the reference cell but with absolute DEF varying by up to a factor of 3 in some cases.  There is a large change in DEF with cell radius, which is primarily caused by the change in single cell GNP uptake caused by the changing cell number density (see Equation~(\ref{eq:conc})) and the change in absolute scoring volume.  For cells of the same cell radius but with different nucleus radii, there is an additional effect caused by the relative proximity of the scoring volumes to the gold-containing regions (\ie scoring volume in the small nuclei is, on average, closer to the GNPs in the \peri configuration). 

 Outside of the aforementioned effects, there is also a difference in the \cellDEFs{} trends with energy at points above the L- (20 and 30~keV) and K-edges (90 and 100~keV).  For example, cDEF of the (10, 7)~\um cell peaks at 20~keV and for the (10, 9)~\um cDEF peaks at 30~keV. This discrepancy in cDEF trends going from 20~to~30~keV for different nucleus radii in the large cell is caused by several competing effects. The first is the decrease in gold's photoelectric cross-section above the L-edges ($\sim\,$13.3~keV), leading to fewer photoelectrons being generated in gold at 30~keV than 20~keV for the same number of incoming photons. Photoelectrons generated in the two cases will also have different ranges and stopping powers (10 and 20~keV energy electrons have ranges of 2.5 and 8.6~\um in water, respectively\cite{Be98a}). At 30~keV, the lower total number of photoelectrons is often outweighed by the increased energy deposited by the more energetic photoelectrons, leading to an overall increase in DEF when the scoring volume is large enough, as the more energetic (and long range) electrons still deposit most of their energy before leaving the volume (\eg (10, 7)~\um cell).  In cases where the scoring volume is small (\eg (10, 9)~\um cell with its 1~\um thick cytoplasm shell), the more energetic electrons escape before depositing all of their energy so the increased energy of the photoelectrons does not compensate for the lower number of total photoelectrons, leading to a drop in DEF going from 20 to 30~keV.  This difference in \cellDEFs{} trends is notable for several different cell/nucleus sizes when going from 20 to 30~keV and 90 to 100~keV.

Collectively, this work calculates over \num[group-separator={,}]{4980} unique DEFs (\num[group-separator={,}]{13032} when including the 13 cell cluster simulations in Part~II), of which \num[group-separator={,}]{768} are presented herein. Though this work explores a wide range of parameters, and quantifies many individual trends, perhaps the most important take-away is the potential factor of 21 variation in DEF observed in Figure~\ref{fig:SenseWhisker} depending on cell size, photon energy, and GNP configuration. While some of the individual parameters in this work have been explored in previous studies\cite{Do13,KG15,Su17,RM19,Ca13,Le17}, this is the first work to combine them in such a way as to achieve a more comprehensive view of the factors affecting \cellDEFs{}s in GNPT. The effects of cell/nucleus size and GNP configuration on \cellDEFs{}s cannot be understated, and any GNPT treatment plans need to account for the expected distribution of GNPs in the cell as well as the expected shape of the cell.  Beyond prediction, the cell and GNP configuration data can be used to select optimal beam qualities (\eg orthovoltage units or brachytherapy seeds) to improve GNPT treatment plans. In addition to the single cell work seen here, the cell models and DEF data are used in Part~II to calculate \cellDEFs{}s over macroscopic tumor volumes.

\section{Conclusion} \label{sec_Cell:Conclusion}

Cellular DEFs vary greatly with cell size, gold concentration, and incident source energy.  The \cellDEFs{}s calculated using the lattice model with discrete GNPs are the most realistic representation, though the contiguous volume of pure gold or homogeneous gold-tissue mixture model may be adequate in some scenarios.  DEFs can range from unity to as high as 21.5 and 5.5 for nDEF and cDEF, respectively.  Thus, when proposing GNPT treatments, consideration of many factors beyond those typical to radiotherapy is required, such as the target cell size and the intracellular GNP configuration achieved by the proposed delivery method.  Comparing perinuclear and endosome gold configurations, the large surface area and proximity to scoring volumes of GNPs in the perinuclear configuration led to the largest DEFs for both the cytoplasm and nucleus at all energies and concentrations.  The scaling of DEF with change in gold concentration is linear, but nucleus DEF variation with cell/nucleus size, incoming photon energy, and GNP configuration varies dramatically, from a factor of over 20 in ideal scenarios (small cells, high GNP uptake) to near unity in others (large cells with relatively small nuclei).  The results of this work suggest that,  for GNPT treatment planning,  all treatment parameters (cell/nucleus size, photon energy, and GNP configuration) investigated herein should be considered for both dose estimation and treatment planning purposes, and that a single DEF value that only accounts for GNP concentration is insufficient.

\section*{Acknowledgments}

The authors acknowledge support from the Kiwanis Club of Ottawa Medical Foundation and Dr. Kanta Marwah Scholarship in Medical Physics, the Natural Sciences and Engineering Research Council of Canada (NSERC) [funding reference number 06267-2016], Canada Research Chairs (CRC) program, an Early Researcher Award from the Ministry of Research and Innovation of Ontario, the Queen Elizabeth II Graduate Scholarships in Science and Technology, the Ontario Graduate Scholarship, and the Carleton University Research Office, as well as computing support provided by the Shared Hierarchical Academic Research Computing Network (SHARCNET, \href{https://www.sharcnet.ca}{sharcnet.ca}) and the Digital Research Alliance of Canada (\href{https://www.alliancecan.ca}{alliancecan.ca}). 

\section*{Conflict of interest}

The authors declare that they do not have any conflict of interest.


\setlength{\baselineskip}{0.4cm}

\bibliography{GNP_Cell_Paper_final.bib}

\bibliographystyle{medphy.bst}

\end{document}